\documentclass[conference]{IEEEtran}
\IEEEoverridecommandlockouts
\makeatletter
\def\ps@headings{%
\def\@oddhead{\mbox{}\scriptsize\rightmark \hfil \thepage}%
\def\@evenhead{\scriptsize\thepage \hfil \leftmark\mbox{}}%
\def\@oddfoot{}%
\def\@evenfoot{}}
\makeatother
\usepackage[T1]{fontenc, url}
\usepackage[utf8]{inputenc}
\usepackage{subfigure}
\usepackage{soul}
\usepackage{array}
\usepackage{cite}
\usepackage{amsmath,amssymb,amsfonts}
\usepackage{algorithmic}
\usepackage{graphicx}
\usepackage{textcomp}
\usepackage{xcolor}
\def\BibTeX{{\rm B\kern-.05em{\sc i\kern-.025em b}\kern-.08em
    T\kern-.1667em\lower.7ex\hbox{E}\kern-.125emX}}

\usepackage{algorithmic}
\usepackage[ruled,vlined]{algorithm2e}

\usepackage{url}
\usepackage{pdfpages}

\makeatletter
\def\blfootnote{\xdef\@thefnmark{}\@footnotetext}
\makeatother

\begin{document}
\null%
\includepdf[pages={1}]{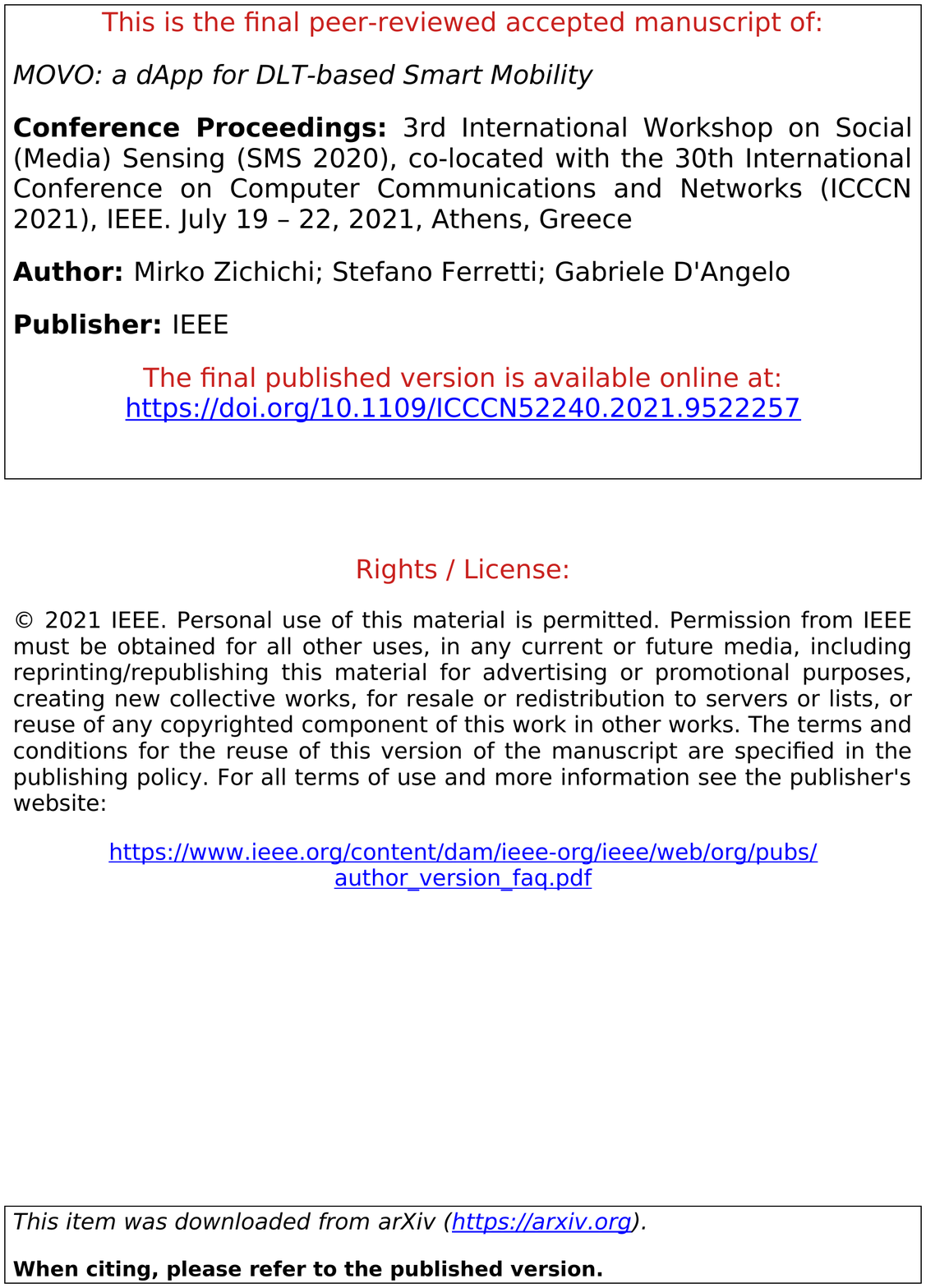}

\title{MOVO: a dApp for DLT-based Smart Mobility
\thanks{This work has received funding from the European Union’s Horizon 2020 research and innovation programme under the Marie Skłodowska-Curie International Training Network European Joint Doctorate grant agreement No 814177 Law, Science and Technology Joint Doctorate - Rights of Internet of Everything.}}

\author{\IEEEauthorblockN{
Mirko Zichichi\IEEEauthorrefmark{1}\IEEEauthorrefmark{2},
Stefano Ferretti\IEEEauthorrefmark{3},
Gabriele D'Angelo\IEEEauthorrefmark{2}}
\IEEEauthorblockA{\IEEEauthorrefmark{1}Ontology Engineering Group, Universidad Politécnica de Madrid, Spain}
\IEEEauthorblockA{\IEEEauthorrefmark{2}Department of Computer Science and Engineering, University of Bologna, Italy}
\IEEEauthorblockA{\IEEEauthorrefmark{3}Department of Pure and Applied Sciences, University of Urbino ``Carlo Bo", Italy} \emph{mirko.zichichi@upm.es, stefano.ferretti@uniurb.it, g.dangelo@unibo.it} 
}
\maketitle

\begin{abstract}
Plenty of research on smart mobility is currently devoted to the inclusion of novel decentralized software architectures to these systems, due to the inherent advantages in terms of transparency, traceability, trustworthiness. MOVO is a decentralized application (dApp) for smart mobility. It includes: (i) a module for collecting data from vehicles and smartphones sensors; (ii) a component for interacting with Distributed Ledger Technologies (DLT) and Decentralized File Storages (DFS), for storing and validating sensor data; (iii) a module for "offline" interaction between devices. The dApp consists of an Android application intended for use inside a vehicle, which helps the user/driver collect contextually generated data (e.g. a driver's stress level, an electric vehicle's battery level), which can then be shared through the use of DLT (i.e., IOTA DLT and Ethereum smart contracts) and DFS (i.e., IPFS). The third module consists of an implementation of a communication channel that, via Wi-Fi Direct, allows two devices to exchange data and payment information with respect to DLT (i.e. cryptocurrency and token) assets. In this paper, we describe the main software components and provide an experimental evaluation that confirms the viability of the MOVO dApp in real mobility scenarios.
\end{abstract}



\section{Introduction}
In the last decade smart mobility has emerged as way to efficiently improve mobility, travel security and increase the options for travellers. The general idea is usually that of devising a sort of middleware to build advanced applications for the provision of innovative transport and traffic management services, with the aim of enabling users ``to be better informed and make safer, more coordinated and ‘smarter’ use of transport networks'' \cite{eu-40-2010}. 

Vehicles and infrastructures are becoming increasingly ``smarter'', which means that they are equipped with sensors that track a huge amount of different types of information, e.g. data sensed by the interior of the vehicle, the surrounding environment, road conditions, etc. In addition, the growth of smartphones and Internet-of-Things (IoT) devices enables individuals' ubiquitous connectivity and the ability to collect personal information or crowdsensing data. All of this constitute a network of devices that is usually referred as VANET (Vehicular Ad-hoc NETwork)~\cite{5247040,leiding2016}.

In this context, a middleware platform can be designed to share and reuse data, services and computation, simplifying the development of new services and the integration of legacy technologies into new ones. In both cases, the use of Distributed Ledger Technologies (DLT) and related cryptocurrencies (or tokens) constitutes a solution that can foster the provision of decentralised services \cite{yuan2016towards,ccnc2020}.  
DLTs, first proposed with Bitcoin's blockchain \cite{nakamoto2008bitcoin}, have radically changed our view of finance, trust in communication and even renewed the concept of contracts and digital democracy \cite{buterin2013ethereum}. The decentralised computation enabled by the DLTs allows us to create self-managed structures that do not depend on central control, avoiding the possibility of a single point of failure.
The incentives for DLT-based Smart Mobility Systems (SMS) would be such that every time a user provides a resource or service based on a smart contract~\cite{leiding2016,lopez2020multi}, he/she earns tokens; every time he/she accesses an SMS resource or service, tokens are consumed (or transferred) \cite{zichichi2020framework}.

In this paper, we propose an application acting within the SMS framework and exploiting the combination of several DLTs. Movo aims to pave the way for a transparent and self-managed system for SMS users, that is not fully managed by a central authority. The proposed architecture focuses on the user's ability to use DLT-based service types and to use smart contracts to exchange data, services and tokens. 

The technology stack includes: the Ethereum public blockchain~\cite{buterin2013ethereum}, widely used for distributed trusted computation through its smart contracts; the IOTA DLT~\cite{popov2016tangle}, that is designed to operate in the IoT landscape; IPFS~\cite{benet2014ipfs}, that enables the storage of data in a decentralized way; Affdex emotion recognition \cite{mcduff2016affdex}, as an example of technology that collects data enabling security measurements for vehicle's drivers; Wi-Fi Direct is used in order to enable device's direct communication. 
The result consists in a mobile decentralized application (dApp), Movo, that allows a vehicle driver to collect real-time data suited to be shared with third parties (e.g. insurance company) and to interact with devices in the VANET for ``off-chain'' tokens transfers. 
In the paper, we provide a description of the developed software architecture and some results to estimate the amount of generated data, in different use cases scenario. 

The rest of this paper is organized as follows. Section \ref{sec:back} provides a background on the technologies used. Section \ref{sec:arch} presents a description of the DLT-based middleware architecture, while in Section \ref{sec:impl} Movo's implementation is provided. Section \ref{sec:usec} presents three use cases with some results and, finally, Section \ref{sec:concl} provides the concluding remarks.
 

\section{Background}\label{sec:back}
In this section we provide an overview of the technology stack implemented in Movo.

\paragraph{Vehicular Ad-hoc Network (VANET)}
A VANET provides a wireless communication network between moving vehicles in order to share safety information, such as accident prevention, post-accident investigation and traffic jams, or other types of non-safety information, such as traveller information \cite{5247040}.
In VANETs, communication takes place between different components: an Application Unit (AU) is a device inside the vehicle that communicates with the vehicle's On-Board Unit (OBU). These two, in turn, can communicate with Road side units (RSUs), devices that are usually fixed along the side of the road to provide services to drivers.
The types of communication in VANETs can be classified into four types: Intra-vehicle consists of a communication between the OBU and one or more AUs; Vehicle-to-Vehicle (V2V) communication occurs between two or more OBUs, either directly or through multiple devices (multi-hop); Vehicle-to-Infrastructure (V2I) is a communication where a vehicle, through OBUs or AUs, engages a dialogue directly with an RSU; Vehicle-to-broadband-cloud, consists of RSU or AU connection to the infrastructure network allowing OBUs to access the entire VANET or the Internet.

\paragraph{Distributed Ledger Technologies}
Distributed Ledger Technologies (DLT) provide a data ledger distributed among a network of peer nodes, where data are written in the form of transactions. This type of technology: guarantees the verifiability of transactions and access to data, i.e. transparency and immutability; shifts trust from a third party intermediary to a distributed consensus mechanism; allows the possibility of direct interactions and agreements between users.
\begin{enumerate}
    \item \emph{Ethereum Smart Contracts.} Ethereum is a blockchain that integrates a (quasi-)Turing-complete language in the form of smart contracts \cite{buterin2013ethereum}. Ethereum can be seen as a transaction-based machine that started with a "genesis" state and step by step executes transaction scripts to switch to a new state. These contracts are called "smart" because they are driven by business logic executed semi-autonomously in order to move assets (i.e. tokens and cryptocurrencies) as the result of an agreement between two parties.
    Once the smart contract is issued in the blockchain, its instructions are immutable and will be executed deterministically. When the issuer is also sure that the implemented behaviour is correct (e.g. through code review), transactions originating from such a contract do not require the presence of a third party to be considered valid between the two contractors.
    \item \emph{IOTA.} IOTA is a DLT specifically designed for the IoT industry, that is based on the use of a Direct Acyclical Graph (DAG), i.e. the Tangle \cite{popov2016tangle}. In this DAG, the vertices represent transactions and the edges represent transactions approvals. When a new transaction is issued, it must approve two previous transactions and the result is represented by directed edges. This mechanism makes it possible to operate in a feeless environment, unlike many other DLTs implementations. Masked Authenticated Messaging (MAM) is a second layer data communication protocol which adds functionality to emit and access an encrypted data stream over the Tangle, regardless of the size or cost of device \cite{zichichi2020framework}. MAM allows to maintain integrity and confidentiality in a data flow produced by a device. MAM works through Channels, where the publisher and viewers meet. Viewers subscribe to channels to get the data that channel owner publishes.
\end{enumerate}

\paragraph{Decentralized File Storage and IPFS}
In order to overcome the typical scalability and privacy issues of DLTs and cloud services, while maintaining the benefits of decentralization, Decentralized File Storages (DFS) are a potential solution for storing files \cite{zichichi2020efficiency}. In practice, DFS are leveraged for storing data outside the DLT, i.e. ``off-chain'' storage, and offer higher data availability and resilience thanks to data replication. The InterPlanetary File System (IPFS)~\cite{benet2014ipfs} is a DFS that builds a file system over a peer-to-peer network. Files published in the IPFS network take the form of IPFS objects and are retrieved through their digest, i.e.~the result of a hash function applied to the file.

\paragraph{Affdex}
Affdex is an emotion measurement technology able to recognize human emotions based on facial cues or physiological responses \cite{mcduff2016affdex}. It is based on the use of the Facial Action Coding System (FACS) and on four main components: face and facial landmark detection, face texture feature extraction, facial action classification and emotion expression modelling. Convolutional Neural Networks (CNN) and Recurrent Neural Networks (RNN) are used provide real-time estimates of emotions on mobile devices. 

\begin{figure}[t]
    \centering
	\includegraphics[width=.49\textwidth]{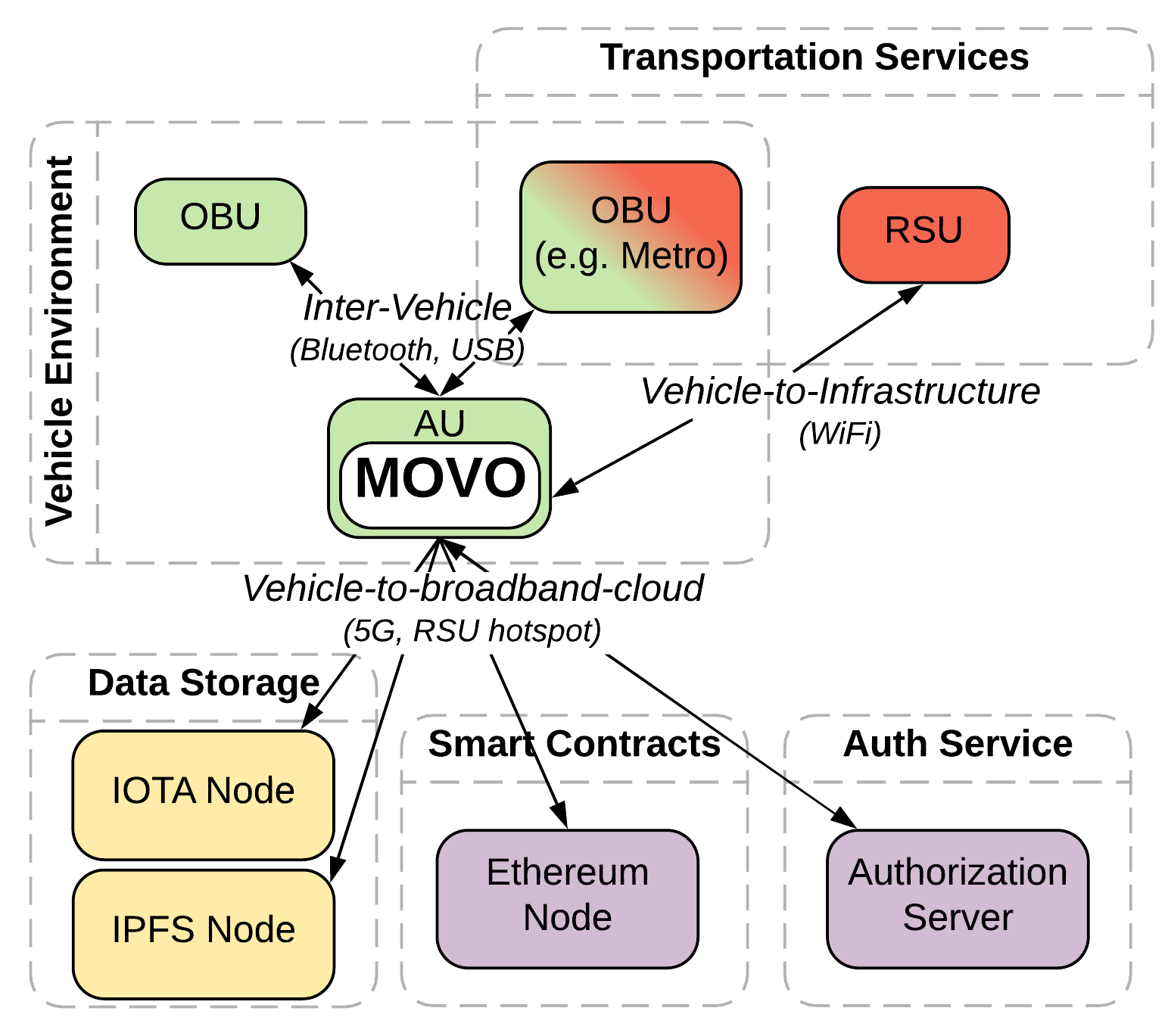}
	\caption{Communication in the DLT-based Middleware Architecture.}
	\label{fig:movocomm}
\end{figure}

\section{DLT-based Middleware Architecture}\label{sec:arch}
The Movo dApp is at the heart of a middleware system built for mobility in SMS, which exploits VANETs and the functionality offered by DLTs. In this middleware system, it is possible to distinguish five different subsystems with which the Movo dApp communicates (Figure \ref{fig:movocomm}):
\begin{itemize}
    \item \emph{Inter-Vehicle.} Movo is an app running in a smartphone, i.e. the AU. An inter-vehicle communication happens between such smartphone and the OBU. The medium for such communication can be the Bluetooth or a USB cable. Actually, Movo is only used to gather data from the OBU, but it could actually also be used to control some OBU's capabilities.
    \item \emph{Vehicle-to-Infrastructure.} Such communication takes place when Movo, through the AU, interacts with the RSUs. In this case, Wi-Fi can be the most appropriate communication medium technology, due the typical distance between the AU and a RSU.
    \item \emph{Vehicle-to-Broadband-Cloud.} The most used type of communication consists in the use of cellular network, i.e. 5G antennas, to access Internet and transmit data, i.e. Vehicle-to-Broadband-Cloud. However, this is not the only case of such category because the AU can directly interact with the RSU to access Internet, e.g. the latter can act as a DLT node.
\end{itemize}
The middleware architecture that we describe in this work provides two main functions: \textit{(a)} the collection and sharing of sensors' data, \textit{(b)} the fruition of DLT-based services (implemented as both on- or off-chain solutions). The main details of these two functions are described in the next two subsection, in isolation.

\paragraph{Collection and Sharing of Sensors' Data}
\begin{enumerate}
    \item \emph{Vehicle Environment.} The middleware has been designed with Movo's application at the very core of all functionalities. This smartphone app is designed to work inside a vehicle, connected to the OBU and directly accessible by the driver. Movo communicates internally with the smartphone, i.e. AU, and the OBU, capturing travel data from multiple sensors and recording safety and comfort events.
    \item \emph{Data Storage.} To guarantee data integrity and verifiability, encrypted sensors data are stored directly in IPFS and the Tangle is used to reference data and their content on-chain, e.g. through an hash pointer. Collected data, thus, are indexed and accessed thanks to the use of MAM Channels. Preventing the on-chain storage is a preferable solution, not only for maintaining higher availability for data reads and better performances for data writes \cite{zichichi2020efficiency}, but also because it is generally incompatible with data protection requirements \cite{finck2019blockchain}. MAM channels are personal for each user and created directly (and automatically) through Movo.
    \item \emph{Smart Contracts for Data Sharing.} The Ethereum blockchain is fundamental to let share and obtain sensors' data through smart contracts. Access to the data stored in IPFS and indexed by MAM channels can be allowed by the data owner indicating the receiver's Ethereum address in a specific smart contract. The release of keys for accessing the encrypted data and MAM Channels, then, is authorized only to entitled users and executed by an authorization service.
    \item \emph{Authorization Service.} The authorization service is designed as a process that enforces the access rights that are specified in the smart contracts. Here the authorization servers are in charge of providing the encryption key of MAM channels and IPFS objects to every requester that can rightfully access it. This operation is preceded by a check on the specific smart contract that contains a list of all authorized users. 
\end{enumerate}

\begin{figure*}[!ht]
    \centering
	\includegraphics[width=.995\textwidth]{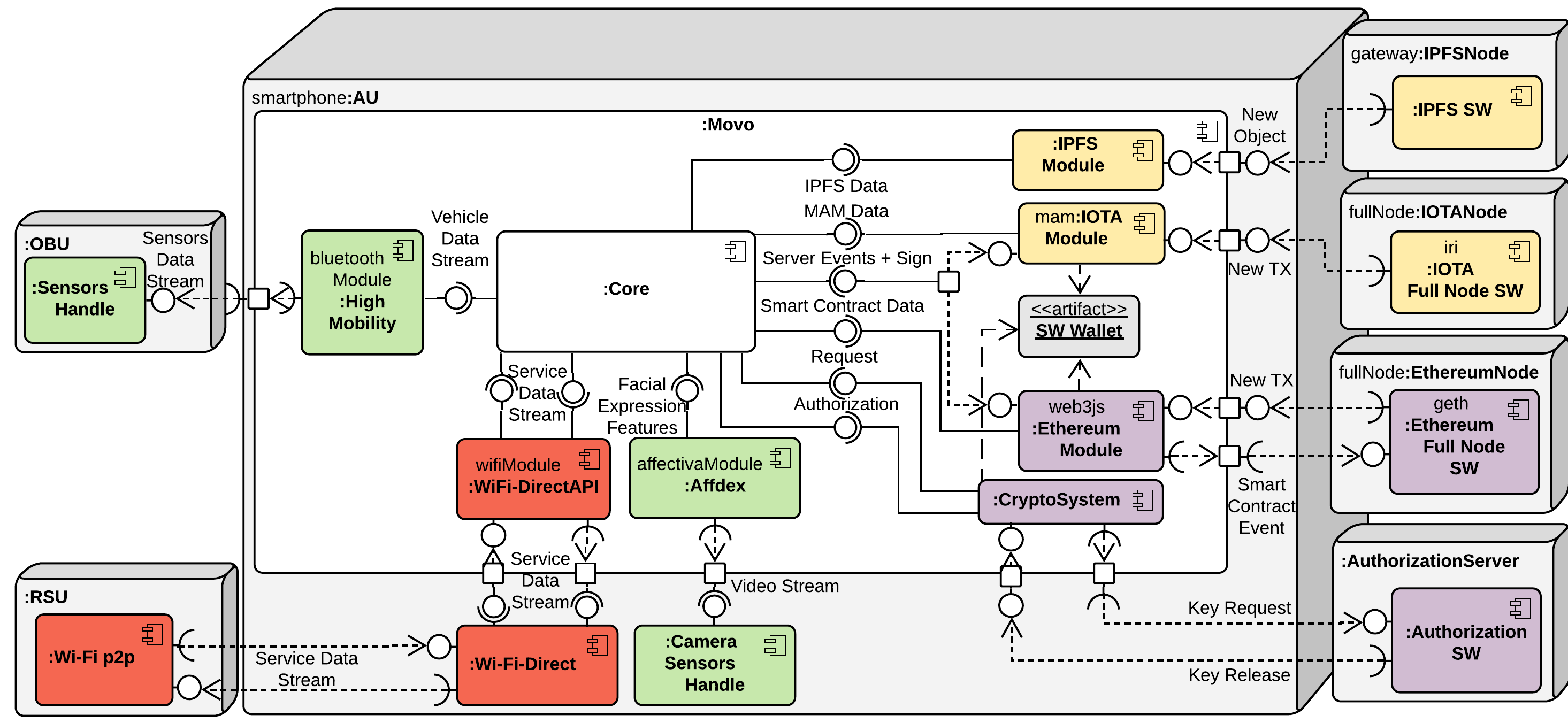}
	\caption{Movo Components Diagram and related middleware's systems. Components in green are part of the Vehicle Environment, red ones belong to the Transportation Services, yellow ones to Data Sharing and purple ones to Smart Contracts and Authentication System.}
	\label{fig:movocomponents}
\end{figure*}

\paragraph{Fruition of DLT-based Services}
\begin{enumerate}
    \item \emph{Transportation Services.}  
    In SMS, services to drivers are supplied by one or more providers that leverage the VANET to communicate (e.g. motorway tolls, electric charging stations).
    RSUs can offer various services to drivers and Movo makes use of those through a direct communication, i.e. through Wi-Fi Direct or Bluetooth. An instance of a service is the issuing of location certificates by 'trusted' RSUs, stating that a driver/user passed by them at a certain time.
    \item \emph{Smart Contracts for Services.} Other than the smart contracts created for data sharing, users and service providers can create ad-hoc smart contracts for the provision of services in SMS. In the middleware architecture, thus, we thought of providing an unified Ethereum ERC20 Token for the payment of such services, i.e. the Movo token. Having a unique token, used and shared among all services, can be beneficial for a global use of all the services offered in the SMS, e.g.~a user that receives a token for having indexed an empty a parking spot, then can make use of it for a car sharing service.
    \item \emph{State channels and Micropayments.} Since transactions in smart contracts and DLTs can be expensive in terms of fees and latencies \cite{zichichi2020are}, state channels have been introduced to implement rapid micropayments. These consist in a design pattern for instant DLT transactions made off-chain, where only the first and the last payment transactions are stored into the ledger. This channel enables the exchange of transactions that update a token/coin balance through an off-chain communication medium. In Movo such medium consists in a direct communication, i.e. through Wi-Fi or Bluetooth.
\end{enumerate}


\section{Movo: Technical Implementation Details}\label{sec:impl}
Movo is a Decentralized Application (dApp) \cite{buterin2013ethereum}, implemented as an Android app, that dialogues with several systems, as described in the middleware architecture related section above. Movo is composed of the following modules (see Figure \ref{fig:movocomponents}): 
\paragraph{Core module}
Movo's core module controls every other module and the information on screen, i.e. the User Interface. This core module mainly acts as:
\begin{enumerate}
    \item \emph{gateway}, for the communication among Movo modules; information coming from the smartphone's sensors is directed to the Core and then routed to the appropriate module for communication with external services.
    \item \emph{user interface}, as this module implements a graphical user interface to show, on the smartphone screen, elements and views related to the whole system.
\end{enumerate}
\paragraph{High Mobility module}
As concerns the inter-vehicles communication, Movo communicates with the OBU. To this aim, we take advantage of High-Mobility \cite{highmobility2019}, a platform 
offering a set of 
standardized APIs for communication with vehicles. Movo receives (and can also possibly send) information via Bluetooth through the High-Mobility module.
\paragraph{Affectiva module}
Affdex library is used to check driver's emotions and facial expressions, e.g. to detect behavior anomalies. In particular, the Affectiva module directly connects to the smartphone camera and starts analyzing 10 frames per seconds of raw footage, returning various emotions and expressions indices. Eyes closure, attentions, anger, surprise and others are decoded into values between 0 and 1 and then associated all the faces appearance measurements framed by the camera. These data are directly uploaded to IPFS and the hash digest issued to the dedicated MAM Channel. 
\paragraph{Wi-Fi Direct module}
Wi-Fi Direct module exploits Android's Wi-Fi Direct libraries to manage peer-to-peer direct communications. Through this module, Movo discovers peers, connects and communicates with them. This module is used for the interaction with the RSUs. Hence, different kind of messages, such as location certificates requests or micropayments receipt pass though this channel. 
\paragraph{Data Sharing modules}
When the core module receives data from sensors, it organizes such data by type (e.g. location, affdex etc.) and time interval (e.g. batch of measurements in the last 10 seconds). Such data are encrypted with a symmetric key. Then, the following processes occur:
\begin{enumerate}
    \item each encrypted data packet is sent to the IPFS module to be uploaded in the IPFS network. 
    \item the digest of the encrypted data packet is sent to the IOTA module to be issued to the relative MAM channel; the address of such channel is stored in an Ethereum smart contract.
    \item the symmetric key is shared with the Authorization Service through the cryptosystem module, using a distributed mechanism as described in \cite{zichichi2020personal}. 
\end{enumerate}
\paragraph{Smart Contracts module and Wallet}
Transactions to create, update, delete or interact with smart contracts or data sharing and SMS services are generated in the Ethereum module. 
It is worth noting that, for the sake of secure communications, all these modules maintain private and symmetric keys in an internal software wallet \cite{fan2019secure}.


\section{Use Case Scenarios and Evaluation}\label{sec:usec}
\begin{figure}[t]
    \centering
	\includegraphics[width=.49\textwidth]{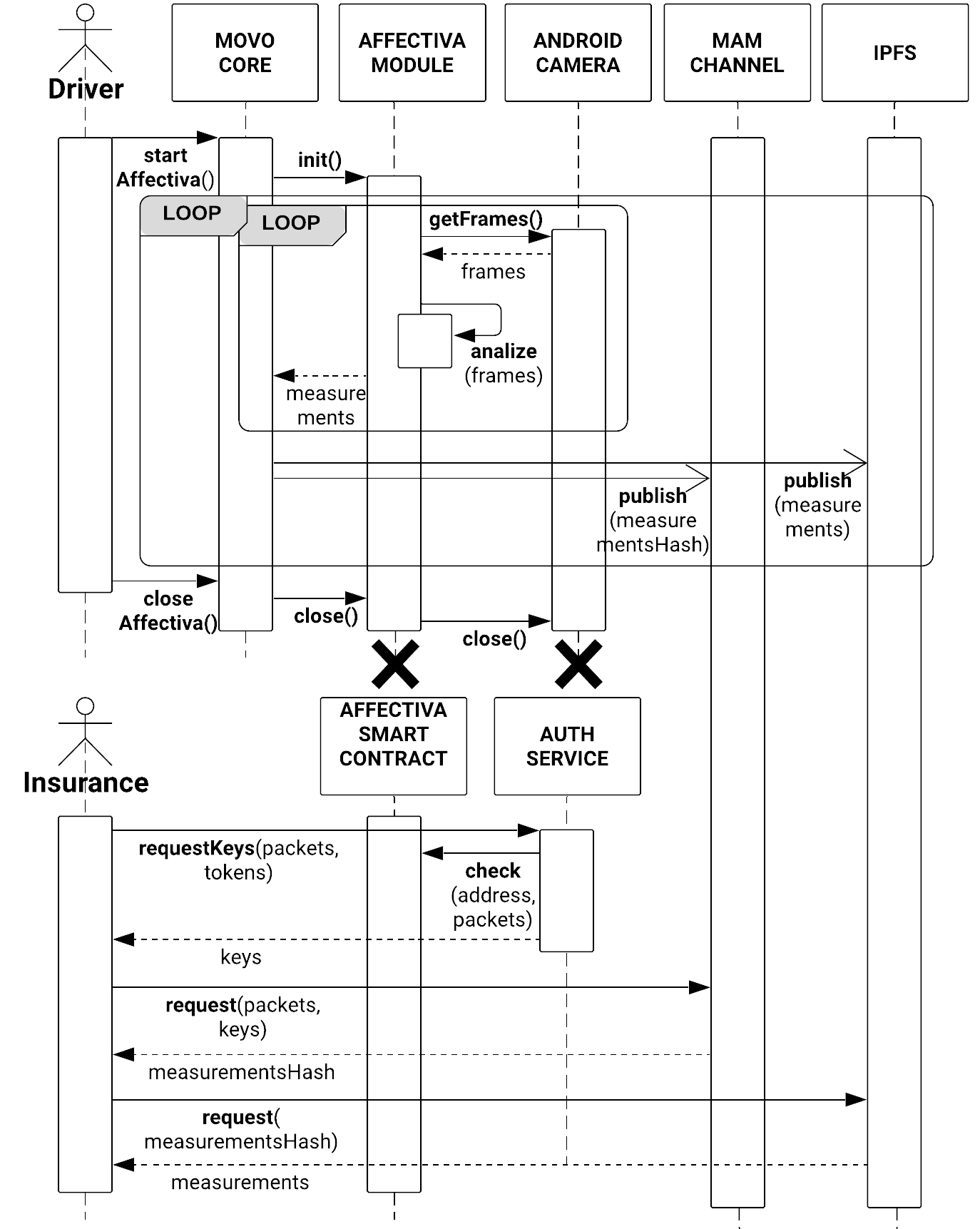}
	\caption{Vehicle insurance monitoring scenario sequence diagram}
	\label{fig:movoseq}
\end{figure}
In this section, we provide three scenarios regarding the use of Movo and the middleware implementation, together with some considerations regarding the performance evaluation. 
\paragraph{Vehicle Insurance Monitoring}
Due to the immutability of the sensor data (in the form of hash digests stored in the DLT) Movo can be successfully implemented to provide accountability in case of a dispute, in the relationship between car owners and insurance companies.
A possible scenario consists in the measurement of driver's emotions and expressions to detect unsuitable driving conditions. The driver (the car owner or someone else) starts the Affectiva module through the Movo GUI and places the smartphone so that the front camera is able to capture her face. In Movo, a loop is triggered (Figure \ref{fig:movoseq}) in which the acquired data, i.e. 10 frames and related measurements, are uploaded each second to IPFS and then their hash digests are grouped and published every 20 sec to the relative MAM channel via the IOTA module. 
Data acquisition consists in an inner loop where the Affectiva module captures the stream from the smartphone's camera (with ten frames per second) and then analyzes it using its classifiers to return expressions and emotions measurements. 
Later, anyone can access these measurements through a request to Ethereum's smart contract and the subsequent release of the keys if the request is accepted. Therefore, since the insurance company address was entered into the smart contract before by the car owner, the former only needs to request the keys of specific data packets to the Authorization Service, which, in turn, checks eligibility before the release.

This scenario requires 1MB/sec to upload camera frames in IPFS (see Table \ref{fig:table}), but most importantly 10 requests per second. If we take, for instance, the average daily travel time of an Italian ($\sim$70 min 
\cite{euro2012}), the result would be $\sim$4.2GB of data to provide to the insurance company per user per day. Furthermore, such data are supplemented by metadata, i.e. affdex measurements, in order to streamline upstream data analysis.

\paragraph{MyMovoMechanic Service}
Another application scenario is related to remote analyses of vehicles components status (MyMovoMechanic). This service can be offered through a dedicated smart contract, where a car owner can register or ``check-in'' and (automatically) pay to obtain car maintenance. The reference to the smart contract address can be directly listed in the car owner's eligible data consumers list. The process is similar to the one shown in Figure \ref{fig:movoseq}; however, in this case the High Mobility module is the one that fetches data, while MyMovoMechanic company can be compared to the Insurance actor. 

To support such a kind of service, a high number of requests per second to IPFS is required (e.g. 90/sec when we simulate a Porsche Cayenne in High Mobility, Table \ref{fig:table}), but with smaller data size w.r.t. the previous use case ($\sim$300Bytes per data point).
In \cite{zichichi2020efficiency}, we observed that the system is able to scale up to a certain point, when we employ limited resources in the DFS. Thus, to properly support such applications, an adequate amount of dedicated DFS nodes should be provided (e.g. specialized RSUs). The same is true for IOTA; in fact, in \cite{zichichi2020are} we experienced an average latency for MAM messages of $\sim$20 sec (in the best case), thus leading us to aggregate the data and send it with larger intervals.

\begin{table*}
\caption{Data sizes and frequencies in the use-case scenarios.}
\begin{center}
 \begin{tabular}{| m{2.5cm} | m{3.7cm}| m{3cm} | m{1.5cm} | m{2.5cm} |}
 \hline
 \textbf{Scenario} & \textbf{Sensor Data} & \textbf{Size} & \textbf{Frequency} & \textbf{TOT} \\ 
 \hline
 \hline
 Insurance Monitoring & Camera frame & $\sim$100KB (640x480px) & 10/sec & $\sim$60MB/min (IPFS) \\ 
 \hline
 Insurance Monitoring & Camera frames hashes & 1 MAM message (3 TXs) & 1/20sec & 3/min (IOTA) \\ 
 \hline
 MyMovoMechanic Service & Vehicle data point & $\sim$300Bytes (256 chars + JSON + timestamp) & 90/sec & $\sim$1.62MB/min (IPFS) \\
 \hline
 MyMovoMechanic Service & Vehicle data point hashes & 1 MAM message (3 TXs) & 1/20sec & 3/min (IOTA) \\ 
 \hline
 Charging Service & Open/Close payment channel & 1 Ethereum TX & 2 per service & 2 TXs (Ethereum) \\ 
 \hline
Charging Service & Off-chain balance update msgs & 1 Wi-Fi Direct msg & Several per session & Several (off-chain)\\ 
 \hline
\end{tabular}
\end{center}
\label{fig:table}
\end{table*}

\paragraph{Electric Car Charging Service}
\begin{figure}[t]
    \centering
	\includegraphics[width=.49\textwidth]{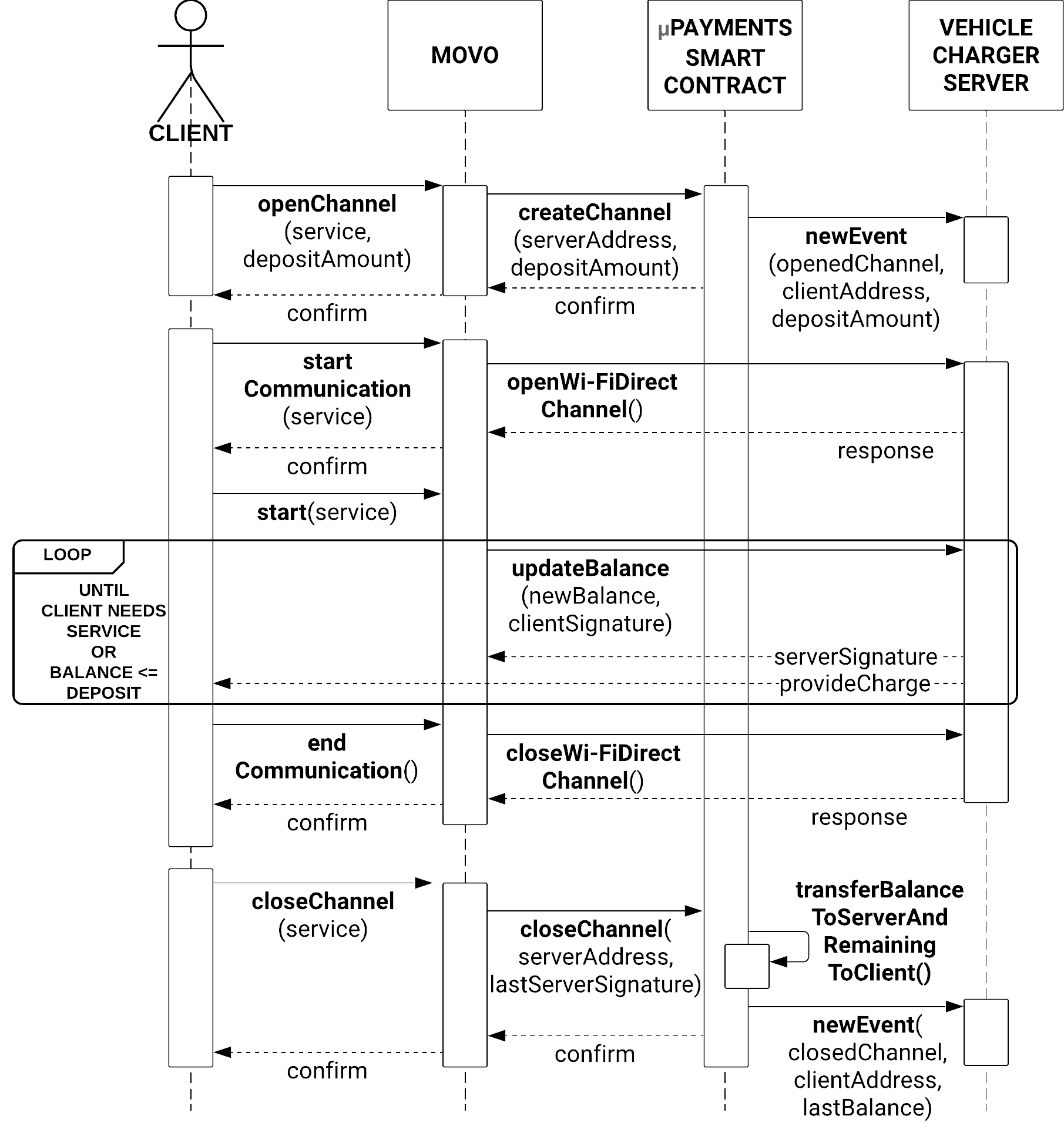}
	\caption{Electric car charging service scenario sequence diagram}
	\label{fig:movoseq2}
\end{figure}
Movo has been designed to communicate directly with devices on the road and in particular to make use of off-chain ``instant'' payments. A further possible scenario, that exploits this feature, consists in the use of charging services for electric vehicles. In this case, the smartphone running Movo and the device providing the service interact through Wi-Fi. We can distinguish three moments that can occur, even far apart in time (Figure \ref{fig:movoseq2}): 
\begin{enumerate}
    \item The opening of the payment channel, which consists in an operation executed by client prior to any interaction with the services provided by the server. This consists in sending a transaction to the dedicated micropayments smart contract, where a deposit is allocated to by the client in favor of the address of the electric vehicle charging company.
    \item The use of the service, in which the client starts a Wi-Fi Direct communication channel with the charger server, in order to: (i) exchange digitally signed messages where a balance (starting from 0) is updated; (ii) receive the service. When the client no longer needs services, the channel can be paused and used again in the future until the balance value reaches the deposit one or until one of the two actors wants to close it. 
    \item The closure of the payment channel (which consists in the second and last on-chain transaction) occurs when one of the two actors submits the final balance to the micropayments smart contract. Then, the smart contract will automatically transfer the balance to the server and the remaining part of the deposit to the client.
\end{enumerate}
In this use case, the amount of data sent to the DLT is very limited (see Table \ref{fig:table}).

\section{Conclusions}\label{sec:concl}
We presented Movo, a DLT-based mobile dApp that allows drivers to manage and distribute data produced by vehicles and users, while on the move. Movo can interact with the vehicle, uses the IOTA Tangle and IPFS to store data in a decentralized way. Moreover, Ethereum smart contracts allows authorized users to get access to these data. 
We provided some results that provide an estimation on the amount of generated data, in different use case scenarios. These results provide insights on the requirements needed at the decentralized systems to adequately support practical and viable mobility applications.
The source code of Movo is available on GitHub: \url{https://github.com/miker83z/movoApp}.

\bibliographystyle{./IEEEtran}
\bibliography{./movo}

\end{document}